\documentclass{ws-procs9x6}

\usepackage{epsfig}

\usepackage{prprntno}

\begin{document}

\title{Effects of CP phases on the Phenomenology of SUSY Particles%
\footnote{\uppercase{I}nvited plenary talk presented by
            \uppercase{A}.~\uppercase{B}artl at 
            \uppercase{PASCOS}'04/\uppercase{N}ath\uppercase{F}est,
            \uppercase{N}ortheastern \uppercase{U}niversity,
            \uppercase{B}oston, \uppercase{MA}, 
            \uppercase{A}ugust 16--22, 2004.}}

\author{A. Bartl and S. Hesselbach}

\address{Institut f\"ur Theoretische Physik, Universit\"at Wien,
  A-1090 Vienna, Austria}

\begin{flushright}
\ttfamily
  UWThPh-2004-34 \\
  hep-ph/0410237
\end{flushright}
\vspace{-6mm}

\maketitle

\abstracts{
We review our recent studies on the effects of CP-violating
supersymmetric (SUSY) parameters on the phenomenology of 
neutralinos, charginos and third generation squarks.
The CP-even branching ratios of the squarks show a
pronounced dependence on the phases of $A_t$, $A_b$, $\mu$ and $M_1$
in a large region of the supersymmetric parameter space,
which can be used to get information on these phases.
In addition we have studied
CP-odd observables, like asymmetries based on triple product
correlations.
In neutralino and chargino production with subsequent three-body
decays these asymmetries can be as large as 20\,\%.}

\section{Introduction}

The Lagrangian of the Minimal Supersymmetric Standard Model (MSSM)
contains several complex parameters, which are new sources of
CP-violation.
In the sfermion sector of the MSSM the trilinear scalar couplings
$A_f$ and the higgsino mass parameter $\mu$ can be complex.
In the chargino and neutralino sector the parameter $\mu$
and the U(1) gaugino mass parameter $M_1$ can be complex, taking $M_2$
real.

The phases of the complex parameters are constrained or
correlated by the experimental upper limits on the electric
dipole moments of electron, neutron and the atoms
${}^{199}$Hg and ${}^{205}$Tl.
In a constrained MSSM the restrictions on the phases can be rather
severe. However, there
may be cancellations between the contributions of different
complex parameters, which allow larger values for the phases
\cite{Ibrahim:2001yv}.

The study of production and decay of 
charginos ($\tilde{\chi}^\pm_i$) and neutralinos ($\tilde{\chi}^0_i$)
and a precise determination of the underlying supersymmetric (SUSY)
parameters $M_1$, $M_2$, $\mu$ and $\tan\beta$
including the phases $\varphi_{M_1}$ and $\varphi_\mu$
will play an important role at the International Linear Collider (ILC)
\cite{LC}.
In \cite{NeuChaParDet} methods to determine these parameters based on
neutralino and chargino mass and cross section measurements have been
presented.
In \cite{Choi:2004rf} the impact of the SUSY phases on chargino,
neutralino and selectron production has been analyzed and
significances for the existence of non-vanishing phases have been defined. 
In \cite{Bartl:2004xy} CP-even azimuthal asymmetries in chargino production 
at the ILC with transversely polarized beams have been
analyzed.

Concerning the determination of the trilinear couplings $A_f$,
detailed studies in the real MSSM have been performed in
\cite{realAf}.
In the complex MSSM the polarization of final top quarks and tau
leptons from the
decays of third generation sfermions can be a sensitive probe of the
CP-violating phases~\cite{Gajdosik:2004ed}.
In~\cite{staupapers} the effects of the CP phases of $A_\tau$, $\mu$
and $M_1$ on production and decay of tau sleptons ($\tilde{\tau}_{1,2}$) and
tau sneutrinos ($\tilde{\nu}_\tau$) have been studied.
The branching ratios of $\tilde{\tau}_{1,2}$ and $\tilde{\nu}_\tau$
can show a strong phase dependence. The expected accuracy in the
determination of $A_\tau$ has been estimated to be of the order of
10\,\% by a global
fit of measured masses, branching ratios and production cross sections.
The impact of the SUSY phases on the decays of the
third generation squarks has been discussed in
\cite{Bartl:2003yp,squarkpapers,Bartl:2003pd} and will be
reviewed in Sec.~\ref{secSquarkdecay}.

In order to unambiguously establish CP violation in supersymmetry,
including the signs of the phases, a measurement of CP-odd
observables is inevitable.
T-odd triple product correlations between momenta and spins of the
involved particles allow the construction of CP-odd asymmetries already
at tree level \cite{Choi:1999cc,tripleproducts}.
Asymmetries of this kind in scalar fermion decays
have been discussed in \cite{ATsfermion}.
T-odd asymmetries in neutralino and chargino production with
subsequent two-body decays have been analyzed in \cite{AT2body}.
For leptonic two-body decays asymmetries up to 30\,\% can occur.
CP-odd observables involving
the polarization of final $\tau$ leptons from two-body decays of neutralinos 
have been studied in \cite{Ataupol}.
T-odd asymmetries in neutralino and chargino production with
subsequent three-body decays \cite{Bartl:2004jj,charginopaper}
will be discussed in Sec.~\ref{SecAT}.

\section{\label{secSquarkdecay} Decays of Third Generation Squarks}

In \cite{Bartl:2003yp,squarkpapers,Bartl:2003pd} we have studied the effects of
the phases of
the parameters $A_t$, $A_b$, $\mu$ and $M_1$ on the phenomenology of the
third generation squarks, the top squarks
$\tilde{t}_{1,2}$ and the bottom squarks
$\tilde{b}_{1,2}$.
We have focused especially on the effects of $\varphi_{A_t}$ and
$\varphi_{A_b}$ in order to find methods to determine these
parameters.
The third generation squark sector is particularly interesting because
of the effects of the large Yukawa couplings.
The phases of $A_f$ and $\mu$ enter directly the squark mass
matrices and the squark-Higgs couplings, which can cause a strong
phase dependence of suitable observables.
In the case of top squarks the $\mu$ term 
in the off-diagonal element of the mass matrix
is suppressed by $1/\tan\beta$,
hence the phase dependence of the decay matrix elements is essentially
determined by $\varphi_{A_t}$
in a large part of the SUSY parameter space with $|A_t| \gg |\mu|/\tan\beta$.
This can lead to a strong phase dependence of many
partial decay widths and branching ratios.

In the case of bottom squarks the mixing is smaller because of the
smaller bottom quark mass. It is only important for large
$\tan\beta$, when the $\mu$ term in the off-diagonal element of the
mass matrix is large.
However, in the squark-Higgs couplings
the phase $\varphi_{A_b}$ appears independent of the bottom squark mixing.
This can lead to a strong $\varphi_{A_b}$ dependence of bottom squark
\emph{and} top squark partial decay widths into Higgs bosons.

We first discuss the $\varphi_{A_t}$ and $\varphi_{A_b}$
dependence of top squark and bottom squark partial decay widths and branching
ratios. We have analyzed fermionic decays 
$\tilde{q}_i \to \tilde{\chi}^\pm_j q'$,
  $\tilde{q}_i \to \tilde{\chi}^0_j q$
and bosonic decays
$\tilde{q}_i \to \tilde{q}_j' H^\pm$,
  $\tilde{q}_i \to \tilde{q}_j' W^\pm$,
  $\tilde{q}_2 \to \tilde{q}_1 H_i$,
  $\tilde{q}_2 \to \tilde{q}_1 Z$
of $\tilde{t}_{1,2}$ and $\tilde{b}_{1,2}$.
In the complex MSSM the CP-even and CP-odd neutral Higgs bosons mix
and form three mass eigenstates $H_{1,2,3}$ \cite{CPhiggs}.
This can also cause a phase dependence of the widths of the decays into
Higgs bosons.

\begin{figure}[t]
\centerline{\epsfig{file=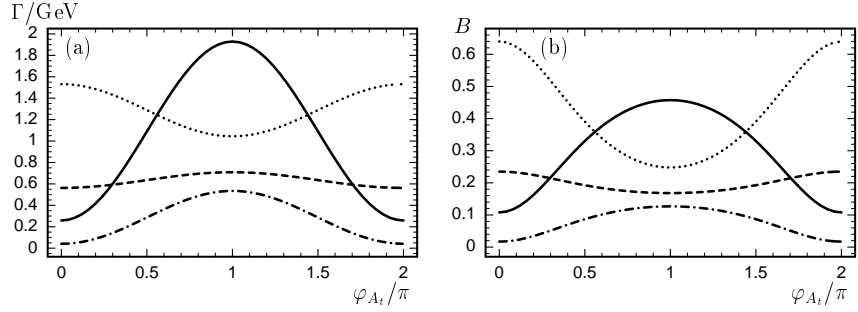,scale=0.71}}
\caption{\label{fig:stop1decays}
(a) Partial decay widths $\Gamma$
and (b) branching ratios $B$ of the decays
$\tilde{t}_1 \to \tilde{\chi}^+_1 b$ (solid),
$\tilde{t}_1 \to \tilde{\chi}^0_1 t$ (dashed),
$\tilde{t}_1 \to \tilde{\chi}^+_2 b$ (dashdotted) and
$\tilde{t}_1 \to \tilde{\chi}^0_2 t$ (dotted)
for
$\tan\beta = 50$, $M_2=233.2$~GeV,
$|M_1|/M_2 = 5/3 \, \tan^2\theta_W$, $|\mu|=377.0$~GeV,
$|A_t|=498.9$~GeV,
$\varphi_\mu=\varphi_{M_1}=\varphi_{A_b}=0$,
$m_{\tilde{t}_1}=530.6$~GeV, $m_{\tilde{t}_2}=695.9$~GeV,
$m_{\tilde{b}_1}=606.9$~GeV, $M_{\tilde{Q}}>M_{\tilde{U}}$
and $m_{H^\pm}=416.3$~GeV.
From \protect\cite{Bartl:2003yp}.}
\end{figure}

In Fig.~\ref{fig:stop1decays} we show the partial decay widths
$\Gamma$ and branching ratios $B$ of the $\tilde{t}_1$.
All partial decay widths,
especially $\Gamma(\tilde{t}_1 \to \tilde{\chi}^+_1 b)$, 
have a pronounced $\varphi_{A_t}$ dependence, which leads to a
strong $\varphi_{A_t}$ dependence of the branching ratios.
This $\varphi_{A_t}$ dependence of the partial decay widths
is caused by that of the top squark mixing matrix
which enters the respective couplings.
In the case of the heavy $\tilde{t}_2$ many decay channels
can be open and can show a strong $\varphi_{A_t}$ dependence.

\begin{figure}[t]
\centerline{\epsfig{file=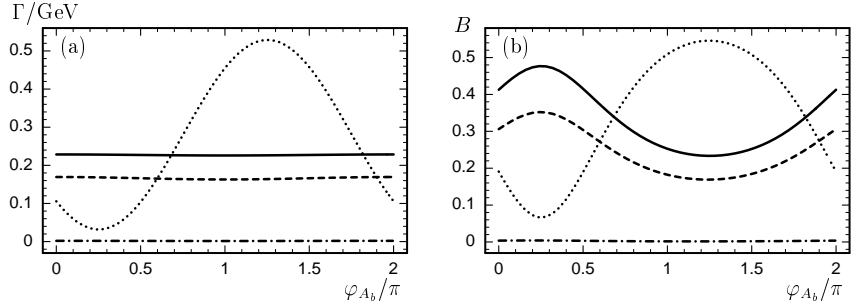,scale=0.7}}
\caption{\label{fig:sbottom1decays}
(a) Partial decay widths $\Gamma$ and (b) branching
ratios $B$ of the decays
$\tilde{b}_1 \to \tilde{\chi}^0_1 b$ (solid),
$\tilde{b}_1 \to \tilde{\chi}^0_2 b$ (dashed),
$\tilde{b}_1 \to H^- \tilde{t}_1$ (dotted) and
$\tilde{b}_1 \to W^- \tilde{t}_1$ (dashdotted)
for $\tan\beta = 30$, $M_2=200$~GeV, 
$|M_1|/M_2 = 5/3 \, \tan^2\theta_W$,
$|\mu| = 300$~GeV, $|A_b|=|A_t|=600$~GeV,
$\varphi_\mu=\pi$, $\varphi_{A_t}=\pi/4$, $\varphi_{M_1}=0$,
$m_{\tilde{b}_1}=350$~GeV, $m_{\tilde{b}_2}=700$~GeV,
$m_{\tilde{t}_1}=170$~GeV,
$M_{\tilde{Q}}>M_{\tilde{D}}$
and $m_{H^\pm}=150$~GeV.
From \protect\cite{Bartl:2003yp}.}
\end{figure}

In Fig.~\ref{fig:sbottom1decays}
we show $\tilde{b}_1$ decay widths and branching ratios.
Only
$\Gamma(\tilde{b}_1 \to H^- \tilde{t}_1)$ shows a pronounced $\varphi_{A_b}$
dependence, which leads to a strong $\varphi_{A_b}$ dependence of the
branching ratios. This is caused by the $\varphi_{A_b}$ 
dependence of the 
$H^\pm \tilde{t}_L \tilde{b}_R$ coupling.
The other partial decay widths depend only very weakly on
$\varphi_{A_b}$.
This is typical for the $\varphi_{A_b}$ dependence of the $\tilde{b}_{1,2}$
decays.
Only the partial decay widths into Higgs bosons,
$\Gamma(\tilde{b}_1 \to H^- \tilde{t}_1)$ for $\tilde{b}_1$ and
$\Gamma(\tilde{b}_2 \to H^- \tilde{t}_{1,2})$, 
$\Gamma(\tilde{b}_2 \to H_{1,2,3} \tilde{b}_1)$ for $\tilde{b}_2$,
can show a strong phase dependence for large $\tan\beta$.

In order to estimate the precision which can be expected in the
determination of the underlying SUSY parameters we have made a global
fit of the top squark and bottom squark decay branching ratios as well as
masses and production cross sections in \cite{Bartl:2003pd}.
In order to achieve this the following assumptions have been made:
(i) At the ILC the masses of the charginos, neutralinos and the
lightest Higgs boson can be measured with high precision.
If the masses of the squarks and heavier Higgs bosons are below
500~GeV, they can be measured with an error of $1\,\%$ and 1.5~GeV,
respectively.
(ii) The masses of the squarks and heavier Higgs bosons, which are
heavier than 500~GeV, can be measured at a 2~TeV $e^+ e^-$ collider
like CLIC with an error of $3\,\%$ and $1\,\%$, respectively.
(iii) The gluino mass can be measured at the LHC with an error of
$3\,\%$.
(iv) For the production cross sections 
$\sigma(e^+ e^- \to \tilde{t}_i \bar{\tilde{t}}_j)$ and
$\sigma(e^+ e^- \to \tilde{b}_i \bar{\tilde{b}}_j)$ and
the branching ratios of the $\tilde{t}_i$ and $\tilde{b}_i$ decays
we have taken the statistical errors, which we have doubled to be on
the conservative side.
We have analyzed two scenarios, one with small $\tan\beta = 6$ and one with
large $\tan\beta = 30$.
In both scenarios we have found that $\mathrm{Re}(A_t)$ and
$|\mathrm{Im}(A_t)|$ can be determined with relative errors of 2 -- $3\,\%$.
For $A_b$ the situation is considerably worse because of the weaker
dependence of the observables on this parameter. The
corresponding errors are of the order of 50\,\%.
For the squark mass parameters
$M_{\tilde{Q}}$, $M_{\tilde{U}}$, $M_{\tilde{D}}$
the relative errors are of order of 1\,\%, for $\tan\beta$ of order of
3\,\% and for the other fundamental SUSY parameters of order of 1 -- 2\,\%.
In this analysis we have used the tree-level formulae for the top
squark and bottom squark decay widths \cite{Bartl:2003pd}.
One-loop corrections are discussed in \cite{ibrahimtalk} and the
references therein.

\section{\label{SecAT}T-odd Asymmetries in Neutralino and Chargino
  Production and Decay}

We have studied T-odd asymmetries
in neutralino \cite{Bartl:2004jj} and chargino \cite{charginopaper}
production with subsequent three-body decays
\begin{equation} \label{ncprocess}
e^+ e^- \to \tilde{\chi}_i + \tilde{\chi}_j \to
\tilde{\chi}_i + \tilde{\chi}^0_1 f \bar{f}^{(')},
\end{equation}
where the full
spin correlations between production and decay have to be included
\cite{spincorr}.
Then in the amplitude squared $|T|^2$ of the combined process products like
$i\epsilon_{\mu\nu\rho\sigma}p^\mu_ip^\nu_jp^\rho_kp^\sigma_l$
of the momenta $p^\mu_i$ of the involved particles
appear in those terms which depend on the spin of the decaying neutralino
or chargino. Together with the complex couplings these terms can give
real contributions to suitable observables at tree-level.
Examples are the triple products
$\mathcal{T}_1 = \vec{p}_{e^-}\cdot(\vec{p}_{f}\times\vec{p}_{\bar{f}^{(')}})$
of the initial electron momentum $\vec{p}_{e^-}$ and
the two final fermion momenta $\vec{p}_{f}$ and $\vec{p}_{\bar{f}^{(')}}$
or
$\mathcal{T}_2 = \vec{p}_{e^-}\cdot(\vec{p}_{\tilde{\chi}_j}\times\vec{p}_{f})$
of the initial electron momentum $\vec{p}_{e^-}$, the momentum of the
decaying neutralino or chargino $\vec{p}_{\tilde{\chi}_j}$ and one
final fermion momentum $\vec{p}_{f}$.
With these triple products we define the T-odd asymmetries
\begin{equation}
A_T = \frac{\sigma(\mathcal{T}_i>0) - \sigma(\mathcal{T}_i<0)}%
 {\sigma(\mathcal{T}_i>0) + \sigma(\mathcal{T}_i<0)}
 =
 \frac{\int {\rm sign}(\mathcal{T}_i) |T|^2 d{\rm Lips}}%
 {{\int}|T|^2 d{\rm Lips}},
\end{equation}
where ${\int}|T|^2 d{\rm Lips}$ is
proportional to the cross section $\sigma$ of the process (\ref{ncprocess}).
$A_T$ is odd under naive time-reversal operation and hence CP-odd, if
higher order
final-state interactions and finite-widths effects can be neglected.

\begin{figure}[t]
\centerline{\epsfig{file=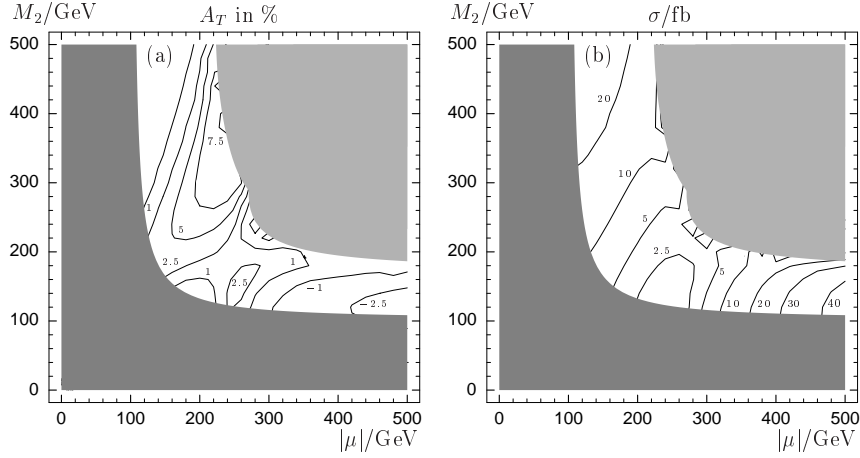,scale=0.71}}
\caption{\label{fig:At12}
Contours (a) of the T-odd asymmetry $A_T$ in \%
and (b) of the cross section
$\sigma(e^+e^- \to \tilde{\chi}^0_1\tilde{\chi}^0_2 \to$
$\tilde{\chi}^0_1\tilde{\chi}^0_1 \ell^+ \ell^-)$,
summed over $\ell=e,\mu\,$,
in fb, respectively,
for $\tan\beta = 10$, $m_{\tilde{\ell}_L} = 267.6$~GeV,
$m_{\tilde{\ell}_R} = 224.4$~GeV, $|M_1|/M_2 = 5/3 \tan^2\theta_W$,
$\varphi_{M_1}=0.5\pi$ and $\varphi_{\mu}=0$
with $\sqrt{s}=500$~GeV and $P_{e^-}=-0.8$, $P_{e^+}=+0.6$.
The dark shaded area marks the parameter space with
$m_{\tilde{\chi}^\pm_1} < 103.5$~GeV excluded by LEP.
In the light shaded area the analyzed three-body decay is strongly
suppressed because $m_{\tilde{\chi}^0_2} > m_Z + m_{\tilde{\chi}^0_1}$
or $m_{\tilde{\chi}^0_2} > m_{\tilde{\ell}_R}$.
From \protect\cite{Bartl:2004jj}.}
\end{figure}

We first consider
neutralino production and subsequent leptonic three-body decay 
$e^+ e^-$ $\to \tilde{\chi}^0_i + \tilde{\chi}^0_2 \to
\tilde{\chi}^0_i + \tilde{\chi}^0_1 \ell^+ \ell^-$
and define the triple product 
$\mathcal{T}_1 = \vec{p}_{e^-}\cdot(\vec{p}_{\ell^+}\times\vec{p}_{\ell^-})$
and the corresponding asymmetry $A_T$.
Then $A_T$ can be directly measured without
reconstruction of the momentum of the decaying neutralino.
We show in Fig.~\ref{fig:At12}
the asymmetry $A_T$ and the corresponding
cross section $\sigma$
for $\tilde{\chi}^0_1 \tilde{\chi}^0_2$ production and subsequent decay of
$\tilde{\chi}^0_2$,
$e^+ e^- \to \tilde{\chi}^0_1 + \tilde{\chi}^0_2 \to
\tilde{\chi}^0_1 + \tilde{\chi}^0_1 \ell^+ \ell^-$.
As can be seen, asymmetries $A_T$ of the order of 10\,\%
can be reached in the parameter region where the cross section is of
the order of 10~fb.
Also for the associated production and decay of $\tilde{\chi}^0_2$ and
$\tilde{\chi}^0_3$,
$e^+ e^- \to \tilde{\chi}^0_3 + \tilde{\chi}^0_2 \to
\tilde{\chi}^0_3 + \tilde{\chi}^0_1 \ell^+ \ell^-$,
the asymmetry $A_T$ has values $\mathcal{O}(10\,\%)$ in large
parameter regions, where the corresponding cross sections 
$\sigma$ are of the order of 10~fb.
For
$e^+ e^- \to \tilde{\chi}^0_4 + \tilde{\chi}^0_2 \to
\tilde{\chi}^0_4 + \tilde{\chi}^0_1 \ell^+ \ell^-$
we have obtained asymmetries $A_T \approx 6\,\%$, however
the cross section is only $\sigma \lesssim 1~\mathrm{fb}$.

\begin{figure}[t]
\centerline{\epsfig{file=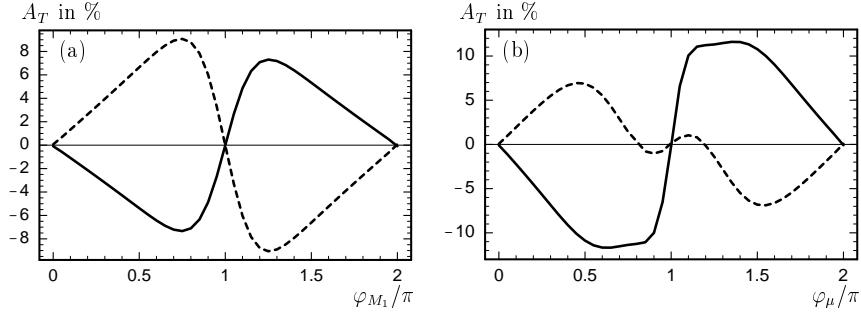,scale=0.71}}
\caption{\label{fig:Atchar12}
T-odd asymmetry $A_T$
for $e^+e^- \to \tilde{\chi}^-_2 + \tilde{\chi}^+_1
 \to \tilde{\chi}^-_2 + \tilde{\chi}^0_1 \bar{s} c$
in the scenario
$M_2=150$~GeV, $|M_1|/M_2= 5/3 \tan^2\theta_W$, $|\mu|=320$~GeV,
$\tan\beta=5$, $m_{\tilde{\nu}}=250$~GeV and $m_{\tilde{u}_L}=500$~GeV
with $\sqrt{s} = 500$~GeV
(a) for $\varphi_{\mu}=0$ and (b) for $\varphi_{M_1}=0$ and
beam polarizations
$P_{e^-}=-0.8$, $P_{e^+}=+0.6$ (solid),
$P_{e^-}=+0.8$, $P_{e^+}=-0.6$ (dashed).}
\end{figure}

We have also studied 
chargino production and subsequent hadronic three-body decay 
$e^+ e^-$ $\to \tilde{\chi}^-_i + \tilde{\chi}^+_1 \to
\tilde{\chi}^-_i + \tilde{\chi}^0_1 \bar{s} c$.
As an example we consider the triple product 
$\mathcal{T}_1 = \vec{p}_{e^-}\cdot(\vec{p}_{\bar{s}}\times\vec{p}_{c})$
and the corresponding asymmetry $A_T$.
In this case it is important to tag the $c$ jet to discriminate
between the two jets and to measure the sign of $\mathcal{T}_1$.
For the associated production and decay of $\tilde{\chi}^+_1$ and
$\tilde{\chi}^-_2$,
$e^+ e^-$ $\to \tilde{\chi}^-_2 + \tilde{\chi}^+_1 \to
\tilde{\chi}^-_2 + \tilde{\chi}^0_1 \bar{s} c$,
asymmetries $A_T$ of the order of 10\,\% are possible
(Fig.~\ref{fig:Atchar12}).
In the scenario of Fig.~\ref{fig:Atchar12} the corresponding cross
sections are in the range of 1 -- 5~fb.
In Fig.~\ref{fig:Atchar12} (b) it is remarkable that large asymmetries
$A_T \approx 10\,\%$ are reached for almost real parameter $\mu$
around $\varphi_\mu = \pi$.
In the chargino sector even for the pair production and decay process
$e^+ e^-$ $\to \tilde{\chi}^-_1 + \tilde{\chi}^+_1 \to
\tilde{\chi}^-_1 + \tilde{\chi}^0_1 \bar{s} c$
asymmetries $A_T \approx 5\,\%$ can appear, which can only originate
from the decay process, because all couplings in the production
process are real.
This means that the contributions from the
decay to $A_T$ play an important role in chargino production with
subsequent hadronic decays.
This can also be seen in 
Fig.~\ref{fig:Atchar12} (a), where $A_T$ can be large for 
$\varphi_\mu = 0$ and $\varphi_{M_1} \neq 0$.
It is furthermore remarkable that 
$\sigma(e^+ e^-$ $\to \tilde{\chi}^-_1 + \tilde{\chi}^+_1 \to
\tilde{\chi}^-_1 + \tilde{\chi}^0_1 \bar{s} c)$
can be rather large, for example 117~fb in the scenario $M_2 = 350$~GeV, 
$|\mu| = 260$~GeV and the other parameters as in 
Fig.~\ref{fig:Atchar12} (a), where $A_T \approx 4\,\%$.

If the momentum of the decaying chargino $\tilde{\chi}^+_1$ can be
reconstructed, for example with help of information from the decay of the
$\tilde{\chi}^-_i$, 
the process
$e^+e^- \to \tilde{\chi}^-_i + \tilde{\chi}^+_1
 \to \tilde{\chi}^-_i + \tilde{\chi}^0_1 \ell^+ \nu$
can be analyzed, where the chargino decays leptonically.
Then the triple product 
$\mathcal{T}_2 = \vec{p}_{e^-}\cdot(\vec{p}_{\tilde{\chi}^+_1}\times\vec{p}_{\ell^+})$
can be used to define $A_T$.
For the associated production and decay of $\tilde{\chi}^+_1$ and
$\tilde{\chi}^-_2$,
$e^+ e^-$ $\to \tilde{\chi}^-_2 + \tilde{\chi}^+_1 \to
\tilde{\chi}^-_2 + \tilde{\chi}^0_1 \ell^+ \nu$,
asymmetries $A_T \gtrsim 20\,\%$ can occur.
But in the region with largest asymmetries around
$|\mu| = 320$~GeV and $M_2 = 120$~GeV the cross section is very
small ($\sigma \approx 0.1~\mathrm{fb}$).
However, for decreasing $|\mu|$ the cross section increases and
reaches $\sigma \approx 2$~fb for $|\mu| = 220$~GeV and $M_2 = 120$~GeV.

\section{Conclusions}

Using the CP-violating MSSM as our framework
we have studied the impact of the complex parameters $A_t$, $A_b$,
$\mu$ and $M_1$ on the decays of top squarks and bottom squarks.
In the case of top squark decays all partial decay widths and branching
ratios can have a strong $\varphi_{A_t}$ dependence because of the large
mixing in the top squark sector.
If $\tan\beta$ is large and decay channels into Higgs bosons are open,
top squark and bottom squark branching ratios can show also a strong
$\varphi_{A_b}$ dependence.
This strong phase dependence of CP-even observables like branching
ratios has to be taken into account in SUSY particle searches at
future colliders.
It will affect the determination of the underlying MSSM parameters.
In order to estimate the expected accuracy in the determination of the
MSSM parameters we have made a global fit of masses, branching ratios
and production cross sections in two scenarios with small and large
$\tan\beta$. We have found that $A_t$ can be determined with an error
of 2 -- 3\,\%, whereas the error of $A_b$ is likely to be of the order
of 50\,\%. 
Furthermore $\tan\beta$ can be determined with an error of 3\,\% and
the other fundamental MSSM parameters with errors of 1 -- 2\,\%.

The measurement of CP-odd observables is inevitable
to unambiguously establish CP violation in supersymmetry.
This will allow to determine the phases including their signs.
We have studied T-odd asymmetries in neutralino and chargino
production with subsequent three-body decays. 
These asymmetries are based on
triple product correlations between incoming and outgoing particles.
They appear already at tree-level because of
spin correlations between production and decay.
The T-odd asymmetries can be as large as 20\,\% and will therefore be
an important tool for the search for CP violation in supersymmetry and
the unambiguous determination of the phases of the SUSY parameters.

\section*{Acknowledgments}

A.B.\ is grateful to the organizers of PASCOS'04/NathFest for
creating an inspiring atmosphere at this conference.
This work has been supported by the European Community's Human Potential
Programme under contract HPRN-CT-2000-00149 ``Physics at Colliders''
and by the ``Fonds zur F\"orderung der wissenschaftlichen For\-schung''
of Austria, FWF Project No.~P16592-N02.


\begin{thebibliography}{99}

%\cite{Ibrahim:2001yv}
\bibitem{Ibrahim:2001yv}
T.~Ibrahim and P.~Nath,
%``Large phases and CP violation in SUSY,''
talk at the \emph{9th International Conference on Supersymmetry
and Unification of Fundamental Interactions,} 11--17 June 2001,
Dubna, Russia, arXiv:hep-ph/0107325 and references therein.
%%CITATION = HEP-PH 0107325;%%

\bibitem{LC}
%\cite{Accomando:1997wt}
%\bibitem{Accomando:1997wt}
E.~Accomando {\it et al.}  [ECFA/DESY LC Physics Working Group
                  Collaboration],
%``Physics with e+ e- linear colliders,''
Phys.\ Rept.\  {\bf 299} (1998) 1
[arXiv:hep-ph/9705442];\\
%%CITATION = HEP-PH 9705442;%%
%
%\cite{Abe:2001np}
T.~Abe {\it et al.}  [American Linear Collider Working Group Collaboration],
%``Linear collider physics resource book for Snowmass 2001. 2: Higgs and
%supersymmetry studies,''
in {\it Proc. of the APS/DPF/DPB Summer Study on the Future of
  Particle Physics (Snowmass 2001) } ed. N.~Graf,
arXiv:hep-ex/0106056;\\
%%CITATION = HEP-EX 0106056;%%
%
%\bibitem{Aguilar-Saavedra:2001rg}
J.~A.~Aguilar-Saavedra {\it et al.}  [ECFA/DESY LC Physics Working Group
                  Collaboration],
%``TESLA Technical Design Report Part III: Physics at an e+e-
% Linear Collider,''
arXiv:hep-ph/0106315;\\
%%CITATION = HEP-PH 0106315;%%
%
%\cite{Abe:2001gc}
%\bibitem{Abe:2001gc}
K.~Abe {\it et al.}  [ACFA Linear Collider Working Group Collaboration],
%``Particle physics experiments at JLC,''
arXiv:hep-ph/0109166.
%%CITATION = HEP-PH 0109166;%%

\bibitem{NeuChaParDet}
%\cite{Choi:1998ei}
%\bibitem{Choi:1998ei}
S.~Y.~Choi, A.~Djouadi, H.~S.~Song and P.~M.~Zerwas,
%``Determining SUSY parameters in chargino pair-production in e+ e-
%collisions,''
Eur.\ Phys.\ J.\ C {\bf 8} (1999) 669
[arXiv:hep-ph/9812236];\\
%%CITATION = HEP-PH 9812236;%%
%
%\cite{Kneur:1999nx}
%\bibitem{kneur}
J.~L.~Kneur and G.~Moultaka,
%``Phases in the gaugino sector: Direct reconstruction of the basic  parameters
%and impact on the neutralino pair production,''
Phys.\ Rev.\ D {\bf 61} (2000) 095003
[arXiv:hep-ph/9907360];\\
%%CITATION = HEP-PH 9907360;%%
%
%\cite{Barger:1999tn}
%\bibitem{Barger:1999tn}
V.~Barger, T.~Han, T.~J.~Li and T.~Plehn,
%``Measuring CP violating phases at a future linear collider,''
Phys.\ Lett.\ B {\bf 475} (2000) 342
[arXiv:hep-ph/9907425];\\
%%CITATION = HEP-PH 9907425;%%
%
%\cite{Choi:2000hb}
%\bibitem{Choi:2000hb}
S.~Y.~Choi, M.~Guchait, J.~Kalinowski and P.~M.~Zerwas,
%``Chargino pair production at e+ e- colliders with polarized beams,''
Phys.\ Lett.\ B {\bf 479} (2000) 235
[arXiv:hep-ph/0001175];\\
%%CITATION = HEP-PH 0001175;%%
%
%\cite{Choi:2000ta}
%\bibitem{Choi:2000ta}
S.~Y.~Choi, A.~Djouadi, M.~Guchait, J.~Kalinowski, H.~S.~Song and P.~M.~Zerwas,
%``Reconstructing the chargino system at e+ e- linear colliders,''
Eur.\ Phys.\ J.\ C {\bf 14} (2000) 535
[arXiv:hep-ph/0002033];\\
%%CITATION = HEP-PH 0002033;%%
%
%\cite{Choi:2001ww}
%\bibitem{Choi:2001ww}
S.~Y.~Choi, J.~Kalinowski, G.~Moortgat-Pick and P.~M.~Zerwas,
%``Analysis of the neutralino system in supersymmetric theories,''
Eur.\ Phys.\ J.\ C {\bf 22} (2001) 563
[Addendum-ibid.\ C {\bf 23} (2002) 769]
[arXiv:hep-ph/0108117, arXiv:hep-ph/0202039];\\
%%CITATION = HEP-PH 0108117;%%
%%CITATION = HEP-PH 0202039;%%
%
%\cite{Gounaris:2002pj}
%\bibitem{Gounaris:2002pj}
G.~J.~Gounaris and C.~Le Mou\"el,
%``The neutralino projector formalism for complex SUSY parameters,''
Phys.\ Rev.\ D {\bf 66} (2002) 055007
[arXiv:hep-ph/0204152];\\
%%CITATION = HEP-PH 0204152;%%
%
%\cite{Choi:2003hm}
%\bibitem{Choi:2003hm}
S.~Y.~Choi,
%``Neutralino pair production and 3-body decays at e+ e- linear colliders as
%probes of CP violation in the neutralino system,''
Phys.\ Rev.\ D {\bf 69} (2004) 096003
[arXiv:hep-ph/0308060];\\
%%CITATION = HEP-PH 0308060;%%
%
%\cite{Choi:2003fs}
%\bibitem{Choi:2003fs}
S.~Y.~Choi and Y.~G.~Kim,
%``Analysis of the neutralino system in two-body decays of neutralinos,''
Phys.\ Rev.\ D {\bf 69}, (2004) 015011
[arXiv:hep-ph/0311037].
%%CITATION = HEP-PH 0311037;%%

%\cite{Choi:2004rf}
\bibitem{Choi:2004rf}
S.~Y.~Choi, M.~Drees and B.~Gaissmaier,
%``Systematic study of the impact of CP-violating phases of the MSSM on
%leptonic high-energy observables,''
Phys.\ Rev.\ D {\bf 70} (2004) 014010
[arXiv:hep-ph/0403054].
%%CITATION = HEP-PH 0403054;%%

%\cite{Bartl:2004xy}
\bibitem{Bartl:2004xy}
A.~Bartl, K.~Hohenwarter-Sodek, T.~Kernreiter and H.~Rud,
%``CP sensitive observables in chargino production with transverse e+- beam
%polarization,''
Eur.\ Phys.\ J.\ C {\bf 36} (2004) 515
[arXiv:hep-ph/0403265].
%%CITATION = HEP-PH 0403265;%%

\bibitem{realAf}
%\cite{Bartl:2000kw}
%\bibitem{Bartl:2000kw}
A.~Bartl, H.~Eberl, S.~Kraml, W.~Majerotto and W.~Porod,
%``Phenomenology of stops, sbottoms, tau sneutrinos, and staus at an e+ e-
%linear collider,''
Eur.\ Phys.\ J.\ directC {\bf 2} (2000) 6
[arXiv:hep-ph/0002115];\\
%%CITATION = HEP-PH 0002115;%%
%
%\cite{Boos:2002wu}
%\bibitem{Boos:2002wu}
E.~Boos, G.~Moortgat-Pick, H.~U.~Martyn, M.~Sachwitz and A.~Vologdin,
%``Impact of tau polarization for the determination of high tan(beta) and
%A(tau),''
arXiv:hep-ph/0211040;\\
%%CITATION = HEP-PH 0211040;%%
%
%\cite{Boos:2003vf}
%\bibitem{Boos:2003vf}
E.~Boos, H.~U.~Martyn, G.~Moortgat-Pick, M.~Sachwitz, A.~Sherstnev and
P.~M.~Zerwas, 
%``Polarisation in sfermion decays: Determining tan(beta) and trilinear
%couplings,''
Eur.\ Phys.\ J.\ C {\bf 30} (2003) 395
[arXiv:hep-ph/0303110].
%%CITATION = HEP-PH 0303110;%%

%\cite{Gajdosik:2004ed}
\bibitem{Gajdosik:2004ed}
T.~Gajdosik, R.~M.~Godbole and S.~Kraml,
%``Fermion polarization in sfermion decays as a probe of CP phases in the
%MSSM,''
arXiv:hep-ph/0405167.
%%CITATION = HEP-PH 0405167;%%

\bibitem{staupapers}
%\cite{Bartl:2002uy}
%\bibitem{Bartl:2002uy}
A.~Bartl, K.~Hidaka, T.~Kernreiter and W.~Porod,
%``Impact of CP phases on the search for sleptons stau and sneutrino/tau,''
Phys.\ Lett.\ B {\bf 538} (2002) 137
[arXiv:hep-ph/0204071];
%%CITATION = HEP-PH 0204071;%%
%
%\cite{Bartl:2002bh}
%\bibitem{Bartl:2002bh}
%A.~Bartl, K.~Hidaka, T.~Kernreiter and W.~Porod,
%``tau-sleptons and tau-sneutrino in the MSSM with complex parameters,''
Phys.\ Rev.\ D {\bf 66} (2002) 115009
[arXiv:hep-ph/0207186].
%%CITATION = HEP-PH 0207186;%%

%\cite{Bartl:2003yp}
\bibitem{Bartl:2003yp}
A.~Bartl, S.~Hesselbach, K.~Hidaka, T.~Kernreiter and W.~Porod,
%``Impact of SUSY CP phases on stop and sbottom decays in the MSSM,''
LC-TH-2003-041, arXiv:hep-ph/0306281;
%%CITATION = HEP-PH 0306281;%%

\bibitem{squarkpapers}
%\cite{Bartl:2003yp}
%\bibitem{Bartl:2003yp}
%A.~Bartl, S.~Hesselbach, K.~Hidaka, T.~Kernreiter and W.~Porod,
%``Impact of SUSY CP phases on stop and sbottom decays in the MSSM,''
%arXiv:hep-ph/0306281;
%%CITATION = HEP-PH 0306281;%%
%
%\cite{Bartl:2003he}
%\bibitem{Bartl:2003he}
A.~Bartl, S.~Hesselbach, K.~Hidaka, T.~Kernreiter and W.~Porod,
%``Impact of CP phases on stop and sbottom searches,''
Phys.\ Lett.\ B {\bf 573} (2003) 153
[arXiv:hep-ph/0307317];
%%CITATION = HEP-PH 0307317;%%
%
%\cite{Bartl:2003pd}
%\bibitem{Bartl:2003pd}
%A.~Bartl, S.~Hesselbach, K.~Hidaka, T.~Kernreiter and W.~Porod,
%``Top squarks and bottom squarks in the MSSM with complex parameters,''
%Phys.\ Rev.\ D {\bf 70} (2004) 035003
%[arXiv:hep-ph/0311338];
%%CITATION = HEP-PH 0311338;%%
%
%\cite{Bartl:2004ws}
%\bibitem{Bartl:2004ws}
%A.~Bartl, S.~Hesselbach, K.~Hidaka, T.~Kernreiter and W.~Porod,
%``Impact of CP phases on the search for top and bottom squarks,''
arXiv:hep-ph/0409347.
%%CITATION = HEP-PH 0409347;%%

%\cite{Bartl:2003pd}
\bibitem{Bartl:2003pd}
A.~Bartl, S.~Hesselbach, K.~Hidaka, T.~Kernreiter and W.~Porod,
%``Top squarks and bottom squarks in the MSSM with complex parameters,''
Phys.\ Rev.\ D {\bf 70} (2004) 035003
[arXiv:hep-ph/0311338];
%%CITATION = HEP-PH 0311338;%%

%\cite{Choi:1999cc}
\bibitem{Choi:1999cc}
S.~Y.~Choi, H.~S.~Song and W.~Y.~Song,
%``CP phases in correlated production and decay of neutralinos in the
%minimal supersymmetric standard model,''
Phys.\ Rev.\ D {\bf 61} (2000) 075004
[arXiv:hep-ph/9907474].
%%CITATION = HEP-PH 9907474;%%

\bibitem{tripleproducts}
%\cite{Donoghue:1977bw}
%\bibitem{Donoghue:1977bw}
J.~F.~Donoghue,
%``T Violation In SU(2) X U(1) Gauge Theories Of Leptons,''
Phys.\ Rev.\ D {\bf 18} (1978) 1632;\\
%%CITATION = PHRVA,D18,1632;%%
%\cite{Kizukuri:1990iy}
%\bibitem{Kizukuri:1990iy}
Y.~Kizukuri and N.~Oshimo,
%``T Odd Asymmetry Mediated By Neutralino In E+ E- Annihilation,''
Phys.\ Lett.\ B {\bf 249} (1990) 449;\\
%%CITATION = PHLTA,B249,449;%%
%\cite{Valencia:1994zi}
%\bibitem{Valencia:1994zi}
G.~Valencia,
%``Constructing CP odd observables,''
arXiv:hep-ph/9411441.
%%CITATION = HEP-PH 9411441;%%

\bibitem{ATsfermion}
%\cite{Bartl:2002hi}
%\bibitem{Bartl:2002hi}
A.~Bartl, T.~Kernreiter and W.~Porod,
%``A CP sensitive asymmetry in the three-body decay stau(1) $\to$  b
%sneutrino/tau tau+,''
Phys.\ Lett.\ B {\bf 538} (2002) 59
[arXiv:hep-ph/0202198];\\
%%CITATION = HEP-PH 0202198;%%
%
%\cite{Bartl:2003ck}
%\bibitem{Bartl:2003ck}
A.~Bartl, H.~Fraas, T.~Kernreiter and O.~Kittel,
%``T-odd correlations in the decay of scalar fermions,''
Eur.\ Phys.\ J.\ C {\bf 33} (2004) 433
[arXiv:hep-ph/0306304];\\
%%CITATION = HEP-PH 0306304;%%
%
%\cite{Bartl:2004jr}
%\bibitem{Bartl:2004jr}
A.~Bartl, E.~Christova, K.~Hohenwarter-Sodek and T.~Kernreiter,
%``Triple product correlations in top squark decays,''
arXiv:hep-ph/0409060.
%%CITATION = HEP-PH 0409060;%%

\bibitem{AT2body}
%\cite{Bartl:2003tr}
%\bibitem{Bartl:2003tr}
A.~Bartl, H.~Fraas, O.~Kittel and W.~Majerotto,
%``CP asymmetries in neutralino production in e+ e- collisions,''
Phys.\ Rev.\ D {\bf 69} (2004) 035007
[arXiv:hep-ph/0308141];
%%CITATION = HEP-PH 0308141;%%
%
%\cite{Bartl:2004ut}
%\bibitem{Bartl:2004ut}
%A.~Bartl, H.~Fraas, O.~Kittel and W.~Majerotto,
%``CP sensitive observables in e+ e- $\to$ neutralino(i) neutralino(j) and
%neutralino decay into Z boson,''
Eur.\ Phys.\ J.\ C {\bf 36} (2004) 233
[arXiv:hep-ph/0402016];
%%CITATION = HEP-PH 0402016;%%
%
%\cite{Bartl:2004vi}
%\bibitem{Bartl:2004vi}
%A.~Bartl, H.~Fraas, O.~Kittel and W.~Majerotto,
%``CP violation in chargino production and decay into sneutrino,''
Phys.\ Lett.\ B {\bf 598} (2004) 76
[arXiv:hep-ph/0406309];\\
%%CITATION = HEP-PH 0406309;%%
%
%\cite{Bartl:2003kn}
%\bibitem{Bartl:2003kn}
A.~Bartl, H.~Fraas, T.~Kernreiter, O.~Kittel and W.~Majerotto,
%``Impact of beam polarization on CP asymmetries in neutralino pair
%production,''
arXiv:hep-ph/0310011;\\
%%CITATION = HEP-PH 0310011;%%
%
%\cite{Kittel:2004kd}
%\bibitem{Kittel:2004kd}
O.~Kittel, A.~Bartl, H.~Fraas and W.~Majerotto,
%``CP sensitive observables in chargino production and decay into a W boson,''
arXiv:hep-ph/0410054.
%%CITATION = HEP-PH 0410054;%%

\bibitem{Ataupol}
%\cite{Bartl:2003gr}
%\bibitem{Bartl:2003gr}
A.~Bartl, T.~Kernreiter and O.~Kittel,
%``A CP asymmetry in e+ e- $\to$ neutralino(i) neutralino(j) $\to$
%neutralino(j)
%tau stau(k) with tau polarization,''
Phys.\ Lett.\ B {\bf 578} (2004) 341
[arXiv:hep-ph/0309340];\\
%%CITATION = HEP-PH 0309340;%%
%
%\cite{Choi:2003pq}
%\bibitem{Choi:2003pq}
S.~Y.~Choi, M.~Drees, B.~Gaissmaier and J.~Song,
%``Analysis of CP violation in neutralino decays to tau sleptons,''
Phys.\ Rev.\ D {\bf 69} (2004) 035008
[arXiv:hep-ph/0310284].
%%CITATION = HEP-PH 0310284;%%

%\cite{Bartl:2004jj}
\bibitem{Bartl:2004jj}
A.~Bartl, H.~Fraas, S.~Hesselbach, K.~Hohenwarter-Sodek and G.~Moortgat-Pick,
%``A T-odd asymmetry in neutralino production and decay,''
JHEP {\bf 0408} (2004) 038
[arXiv:hep-ph/0406190].
%%CITATION = HEP-PH 0406190;%%

\bibitem{charginopaper}
A.~Bartl, H.~Fraas, S.~Hesselbach, K.~Hohenwarter-Sodek and G.~Moortgat-Pick,
in preperation.

\bibitem{CPhiggs}
%\cite{Heinemeyer:2001qd}
%\bibitem{Heinemeyer:2001qd}
S.~Heinemeyer,
%``The Higgs boson sector of the complex MSSM in the Feynman-diagrammatic
%approach,''
Eur.\ Phys.\ J.\ C {\bf 22} (2001) 521
[arXiv:hep-ph/0108059];\\
%%CITATION = HEP-PH 0108059;%%
%
%\cite{Frank:2002qa}
%\bibitem{Frank:2002qa}
M.~Frank, S.~Heinemeyer, W.~Hollik and G.~Weiglein,
%``The Higgs boson masses of the complex MSSM: A complete one-loop
%calculation,''
arXiv:hep-ph/0212037;\\
%%CITATION = HEP-PH 0212037;%%
%
%\cite{Lee:2003nt}
%\bibitem{Lee:2003nt}
J.~S.~Lee, A.~Pilaftsis, M.~Carena, S.~Y.~Choi, M.~Drees, J.~R.~Ellis
and C.~E.~M.~Wagner,
%``CPsuperH: A computational tool for Higgs phenomenology in the minimal
%supersymmetric standard model with explicit CP violation,''
Comput.\ Phys.\ Commun.\  {\bf 156} (2004) 283
[arXiv:hep-ph/0307377].
%%CITATION = HEP-PH 0307377;%%

\bibitem{ibrahimtalk}
T.~Ibrahim, talk at the \emph{10th International Symposium on
Particles, Strings and Cosmology (PASCOS'04/NathFest),}
Northeastern University, Boston, MA, USA, 16--22 August, 2004.

\bibitem{spincorr}
%\cite{Moortgat-Pick:1998sk}
%\bibitem{Moortgat-Pick:1998sk}
G.~Moortgat-Pick, H.~Fraas, A.~Bartl and W.~Majerotto,
%``Spin correlations in production and decay of charginos,''
Eur.\ Phys.\ J.\ C {\bf 7} (1999) 113
[arXiv:hep-ph/9804306];
%%CITATION = HEP-PH 9804306;%%
%
%\cite{Moortgat-Pick:1999di}
%\bibitem{Moortgat-Pick:1999di}
%G.~Moortgat-Pick, H.~Fraas, A.~Bartl and W.~Majerotto,
%``Polarization and spin effects in neutralino production and decay,''
Eur.\ Phys.\ J.\ C {\bf 9} (1999) 521
[Erratum-ibid.\ C {\bf 9} (1999) 549]
[arXiv:hep-ph/9903220].
%%CITATION = HEP-PH 9903220;%%

\end{thebibliography}
\end{document}